\documentclass[aps,prl,twocolumn,showpacs,groupedaddress]{revtex4}

\usepackage{graphicx}
\usepackage{dcolumn}
\usepackage{bm}

\begin{document}



\title{Commensurate-incommensurate solid transition in the $^4$He monolayer 
on $\gamma$-graphyne} 
\author{Jeonghwan Ahn}
\author{Hoonkyung Lee}
\author{Yongkyung Kwon} 
\email{ykwon@konkuk.ac.kr}
\affiliation{
Division of Quantum Phases and Devices, School of Physics,
Konkuk University, Seoul 143-701, Korea
}%

\date{\today}


\begin{abstract}
{Path-integral Monte Carlo calculations have been performed to study the $^4$He adsorption 
on $\gamma$-graphyne, a planar network of benzene rings connected by acetylene bonds. 
Assuming the $^4$He-substrate interaction described by a pairwise sum of empirical 
$^4$He-carbon interatomic potentials, we find that unlike $\alpha$-graphyne, 
a single sheet of $\gamma$-graphyne is not permeable to $^4$He atoms 
in spite of its large surface area.  One-dimensional density
distributions computed as a function of the distance from the graphyne surface 
reveal a layer-by-layer growth of $^4$He atoms. A partially-filled $^4$He monolayer is found to exhibit 
different commensurate solid structures depending on the helium coverage; it shows a C$_{2/3}$ 
commensurate structure at an areal density of 0.0491~\AA$^{-2}$, a C$_{3/3}$ structure 
at 0.0736~\AA$^{-2}$, and a C$_{4/3}$ structure at 0.0982~\AA$^{-2}$.  
While the promotion to the second layer starts beyond the C$_{4/3}$ helium coverage, 
the first $^4$He layer is found to form an incommensurate triangular solid 
when compressed with the development of the second layer.
}
\end{abstract}


\maketitle


For the past few decades, a system of $^4$He atoms adsorbed on a substrate has been 
intensively studied to investigate physical properties of low-dimensional quantum fluids. 
Carbon allotropes have often been chosen as substrates for this purpose because they provide 
strong enough interactions for $^4$He adsorbates to show multiple distinct layered structures~\cite{zimmerli92}. 
As a result of the interplay between $^4$He-$^4$He and $^4$He-substrate interaction, 
these helium adlayers are known to exhibit rich phase diagrams including various commensurate and
incommensurate solids.
On the surface of graphite, a monolayer of $^4$He atoms is crystallized to a C$_{1/3}$ commensurate solid 
at an areal density of $0.0636$~\AA$^{-2}$ and goes through various domain structures 
before freezing into an incommensurate triangular solid 
as the helium coverage increases~\cite{greywall93,crowell96} .
Similar quantum phase transitions were predicted for the $^4$He monolayer 
on a single graphene sheet~\cite{gordillo09,kwon12,happacher13}. 
While no superfluidity has been observed in the first $^4$He layer, the second layer on graphite 
does show finite superfluid response at intermediate helium coverages as first revealed 
by torsional oscillator measurements of Crowell and Reppy~\cite{crowell96}.  
Whether this second-layer superfluid phenomenon is related to two-dimensional supersolidity 
is still an ongoing issue pursued heavily by some experimentalists. 

The $^4$He adsorption on the surface of a carbon allotrope other than graphite or graphene 
has recently been investigated. While $^4$He atoms adsorbed on the interstitials or the groves
of carbon nanotube bundles showed characteristics of one-dimensional quantum fluid~\cite{cole00,gordillo08}, 
a series of theoretical calculations 
predicted well-distinct layered structures
for $^4$He atoms adsorbed on the outer surfaces of fullerene molecules
with each near-spherical helium layer exhibiting various quantum states 
depending on the number of $^4$He adatoms~\cite{kwon10,shin12b,shin13,park14}.  
More recently, graphynes, $sp$-$sp^2$ hybridized two-dimensional networks of carbon atoms~\cite{baughman87,coluci03,coluci04},
have attracted much interest because of their intriguing electronic features such as both symmetric 
and asymmetric Dirac cones~\cite{malko12,kim12} and high carrier mobility~\cite{chen13}. 
Furthermore, they have much larger surface area 
than graphene, which has prompted intensive investigation of their possible applications 
as high-capacity hydrogen storage~\cite{hwang12,koo13} and Li-ion battery anode materials~\cite{hwang13}. 
Using the path-integral Monte Carlo (PIMC) method, one of us recently studied the $^4$He adsorption 
on $\alpha$-graphyne~\cite{kwon13}, a honeycomb structure of both $sp^2$-bonded carbon atoms and $sp$-bonded ones.
Due to the presence of much larger hexagons than those of graphene, 
in-plane adsorption of $^4$He atoms was observed on $\alpha$-graphyne with a single $^4$He atom 
being embedded to the center of each hexagon. The first layer of $^4$He atoms adsorbed on the $^4$He-embedded 
$\alpha$-graphyne was found to undergo a Mott-insulator to commensurate-solid transition
which was interpreted as a transition 
from a spin liquid of frustrated antiferromagnets to a ferromagnetic phase 
with the introduction of Ising pseudospins based on the sublattice symmetry of the honeycomb structure~\cite{kwon13}.

Here we have performed the PIMC simulations to study the $^4$He adsorption on $\gamma$-graphyne, 
the most stable structure among graphynes~\cite{shin14}.
With the increasing number of $^4$He adatoms, 
multiple distinct helium layers are observed on $\gamma$-graphyne. Because of larger hexagons of graphyne, 
these $^4$He adsorbates show a richer phase diagram than the corresponding ones on graphite or graphene. 
Unlike $\alpha$-graphyne, however, even a single sheet of $\gamma$-graphyne is found to be impermeable 
to $^4$He atoms. It is found that the $^4$He monolayer exhibits 
various commensurate solid structures at different areal densities before crystallizing 
into an incommensurate triangular solid at its completion.

\begin{figure}
\includegraphics[width=3.2in]{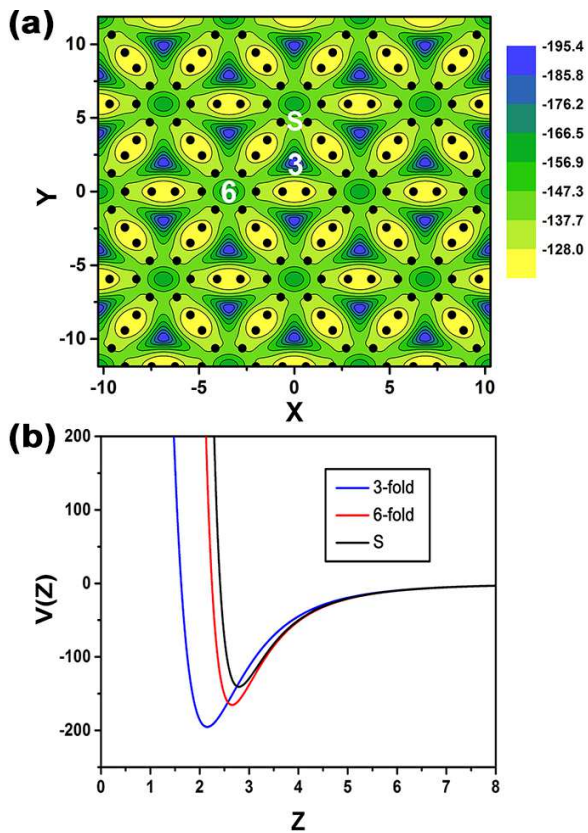}
\caption{(Color online) (a) A contour plot of the minimum $^4$He-graphyne potential, $V_{min}(x,y)$, 
above each point $(x,y)$ on $\gamma$-graphyne and (b) the $^4$He-graphyne potential
as a function of the distance $z$ from the graphyne surface along different symmetry directions. 
The black dots in (a) represent 
the positions of carbon atoms on the $\gamma$-graphyne surface. 
In (b), the blue, the red, and the black line correspond to 
the 3-fold symmetry direction,  the 6-fold direction, and the direction of a saddle point 
(see white numbers and alphabet S in (a)), respectively.
The length unit is~\AA~and the potential energies are in units of Kelvin.
}
\label{fig:pot}
\end{figure} 

In this study, a single $\gamma$-graphyne sheet is fixed at $z=0$ and the helium-graphyne interaction is
described by a sum of pair potentials between the carbon atoms and a $^4$He atom. 
For the $^4$He-C interatomic potential, we use an isotropic 6-12 Lennard-Jones potential proposed 
by Carlos and Cole~\cite{carlos80} 
to fit helium scattering data from graphite surfaces. 
While Fig.~\ref{fig:pot}(a) shows 
a contour plot of the minimum potential energy, 
$V_{min}(x,y)$, above each point $(x,y)$ on the graphyne surface,
Fig.~\ref{fig:pot}(b) presents our $^4$He-graphyne potential 
as a function of the distance $z$ from graphyne along three different symmetry directions 
perpendicular to the graphyne surface.  
As seen in Fig.~\ref{fig:pot}(a), 
there are three adsorption sites per graphyne unit cell, 
two global minima of the $^4$He-graphyne potential located at the centers 
of big irregular hexagons
and one local minimum located at the center of a small regular hexagon (or a benzene ring). 
A bigger hexagon, which has much larger area than a smaller one, is expected
to accommodate more than one $^4$He atom.
Figure~\ref{fig:pot}(b) shows 
that the global minima in the three-fold symmetry directions
are located closer to the graphyne surface by $\sim 0.5$~\AA~than the local minima 
in the six-fold symmetry directions and the potential energy difference between them 
is as large as $\sim 30$ K.
From this we conjecture that the $^4$He adatoms predominantly occupy 
the global minimum sites at low helium coverages, rather than the local minima.  
We here note that there is strong repulsive potential barrier 
for $^4$He atoms as they approach the graphyne surface, {\it i.e.}, $z \rightarrow 0$,
suggesting that $^4$He atoms cannot penetrate through a $\gamma$-graphyne sheet from one side to the other.

This approach of modeling $^4$He-substrate potential with a pairwise sum 
of empirical interatomic potentials has been widely used to study the $^4$He adsorptions 
on various carbon-based substrates, including $\alpha$-graphyne~\cite{kwon13}.
For the $^4$He-$^4$He interaction, we use a well-known Aziz potential~\cite{aziz92}.
Since the exact form of thermal many-body density matrix is not known at a low temperature $T$, 
one can resort to the path-integral representation where the low-temperature density matrix is expressed 
by a convolution of $M$ high-temperature density matrices with an imaginary time step $\tau=1/(Mk_BT)$. 
Both $^4$He-$^4$He and $^4$He-C pair potentials are used to derive the exact two-body density matrices 
at the high temperature $MT$~\cite{ceperley95,zillich05}, which was found to provide accurate description of 
the $^4$He-graphyne interaction as well as the $^4$He-$^4$He interaction with an imaginary time step 
of $(\tau k_B)^{-1}=40$~K. We employ the multilevel Metropolis algorithm described in Ref.~\cite{ceperley95} 
to sample the imaginary time paths as well as the permutations among $^4$He atoms. 
To minimize finite size effects, periodic boundary conditions are applied to 
a fixed $3 \times 2$ rectangular simulation cell
with dimensions of $20.58 \times 23.76$~\AA$^2$. All PIMC simulations presented here started 
from random initial configurations of $^4$He atoms.

\begin{figure}
\includegraphics[width=3.2in]{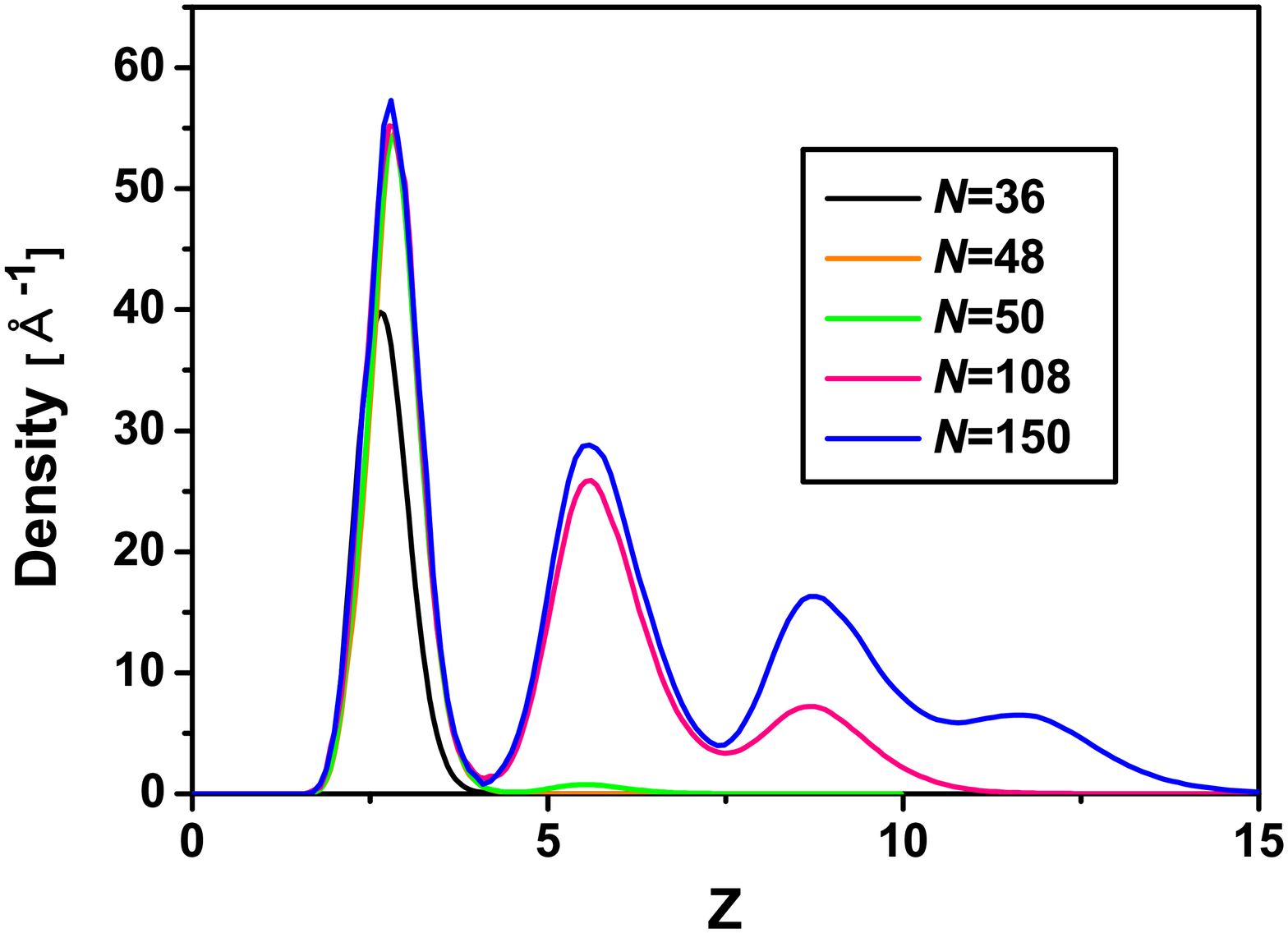}
\caption{(Color online) One-dimensional density of $^4$He atoms adsorbed on a single $\gamma$-graphyne sheet
as a function of the distance $z$ (in \AA) from the graphyne surface. 
Here $N$ represents the number of $^4$He adatoms per $3 \times 2$ rectangular simulation cell 
with dimensions of $20.58 \times 23.76$~\AA$^2$ and the computations were done at a temperature of $0.5$~K. 
} 
\label{fig:1Dden}
\end{figure}

Here we consider the $^4$He adsorption only on one side of the graphyne sheet, {\it i.e.}, $z > 0$.  
Figure~\ref{fig:1Dden} presents one dimensional $^4$He density distributions as a function of distance $z$ 
from the graphyne surface for different numbers of $^4$He adatoms $N$ per $3 \times 2$ rectangular simulation cell.
These density distributions confirm the above assertion 
that $^4$He atoms cannot penetrate to the other side of $z < 0$ through a single graphyne sheet
because of the presence of the strong repulsive potential barrier. 
As more $^4$He atoms are adsorbed, one can see the development of layered structures 
as evidenced by well-distinct density peaks in Fig.~\ref{fig:1Dden}.  
The first sharp peak is located at $z \sim 2.7$~\AA~and the second peak at $z \sim 5.8$~\AA, 
similar to the case of $^4$He on graphene~\cite{kwon12}. We observe the emergence of the $^4$He second layer 
when the number of $^4$He adatoms per $3 \times 2$ simulation cell increases beyond $N=48$ 
(an areal density of $0.0982$~\AA$^{-2}$). With further development of the second helium layer, 
more $^4$He atoms are found to be squeezed into the first layer.
From this we conjecture that the completed first layer would be a compressible incommensurate solid 
like the corresponding layer on graphene~\cite{kwon12} or graphite~\cite{greywall93,corboz08}.  
It is found that the first layer is completed at an areal density of $\sim 0.115$~\AA$^{-2}$ 
while the corresponding value on graphene was predicted to be $\sim 0.12$~\AA$^{-2}$~\cite{kwon12}. 
This small (about 4 \%) difference suggests that the $^4$He-graphene potential 
is more attractive than the $^4$He-graphyne potential 
and graphene may accommodate more $^4$He atoms in its immediate vicinity than graphyne.

\begin{figure}
\includegraphics[width=3.2in]{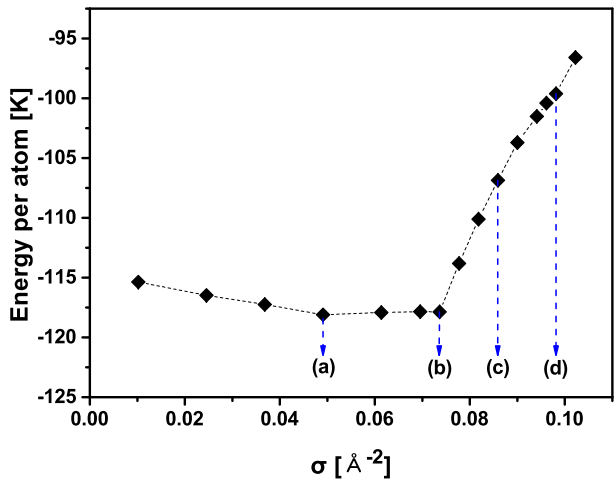}
\caption{(Color online) Energy per $^4$He atom of the first $^4$He layer 
on $\gamma$-graphyne as a function of the helium coverage.
The alphabets (a), (b), (c), and (d) correspond to the areal densities of 0.0491, 0.0736, 0.0859, and 0.0982 ~\AA$^{-2}$, 
respectively, for which two-dimensional $^4$He density plots are shown in Fig.~\ref{fig:2Dden}.
The energies were computed at a temperature of 0.5~K. 
} 
\label{fig:energy}
\end{figure}

Now we discuss the energetics of the $^4$He-graphyne system, 
which provides some insight into the growth of the $^4$He adlayers on $\gamma$-graphyne
and their different quantum phases.
Figure~\ref{fig:energy} shows the energy per $^4$He atom as a function of an areal density $\sigma$. 
At low densities of $\sigma < 0.0736$~\AA$^{-2}$, 
the energy per $^4$He atom changes very little,
indicating that each $^4$He atom occupies one of the adsorption sites, {\it i.e.}, 
the $^4$He-graphyne potential minima.
It is found that the energy per atom has the lowest value at $\sigma=0.0491$ \AA$^{-2}$
which corresponds to two $^4$He atoms per the graphyne unit cell.
Noting that there are two global minima of the $^4$He-graphyne potential per the unit cell 
(see Fig.~\ref{fig:pot}(a)), we conjecture that in the lowest energy state at $\sigma=0.0491$ \AA$^{-2}$
each global minimum site is occupied by a single $^4$He atom.
After filling all global minima,
additional $^4$He atoms are expected to occupy
the local minima located above the centers of the small hexagons,
which is consistent with 
slight increase in the energy per atom for $0.0491$~\AA$^{-2} < \sigma < 0.0736$~\AA$^{-2}$.
Since the distances between the adsorption sites on the graphyne surface
are long enough ($\sim 4$ \AA), the $^4$He-$^4$He interaction is understood to have minimal effects
while $^4$He atoms are filling these adsorption sites.
Each adsorption site, whether it is a global minimum or a local minimum, is occupied by a single $^4$He atom
at an areal density of $\sigma=0.0736$~\AA$^{-2}$, three $^4$He atoms per the graphyne unit cell, 
beyond which one can observe a sudden increase in the energy per atom in Fig.~\ref{fig:energy}.
The continuous increase of the energy per atom for $\sigma > 0.0736$~\AA$^{-2}$ 
suggests that the $^4$He-$^4$He interaction as well as the $^4$He-substrate interaction plays a critical role
in determining quantum states of the $^4$He monolayer at high helium coverages.
One can observe a significant jump in the energy per atom at an areal density of $\sigma=0.0982$~\AA$^{-2}$,
which reflects the start of the second-layer promotion concluded in the analysis of the one-dimensional
density distributions of Fig.~\ref{fig:1Dden}.

\begin{figure*}
\includegraphics[width=6.5in, height=4.0in]{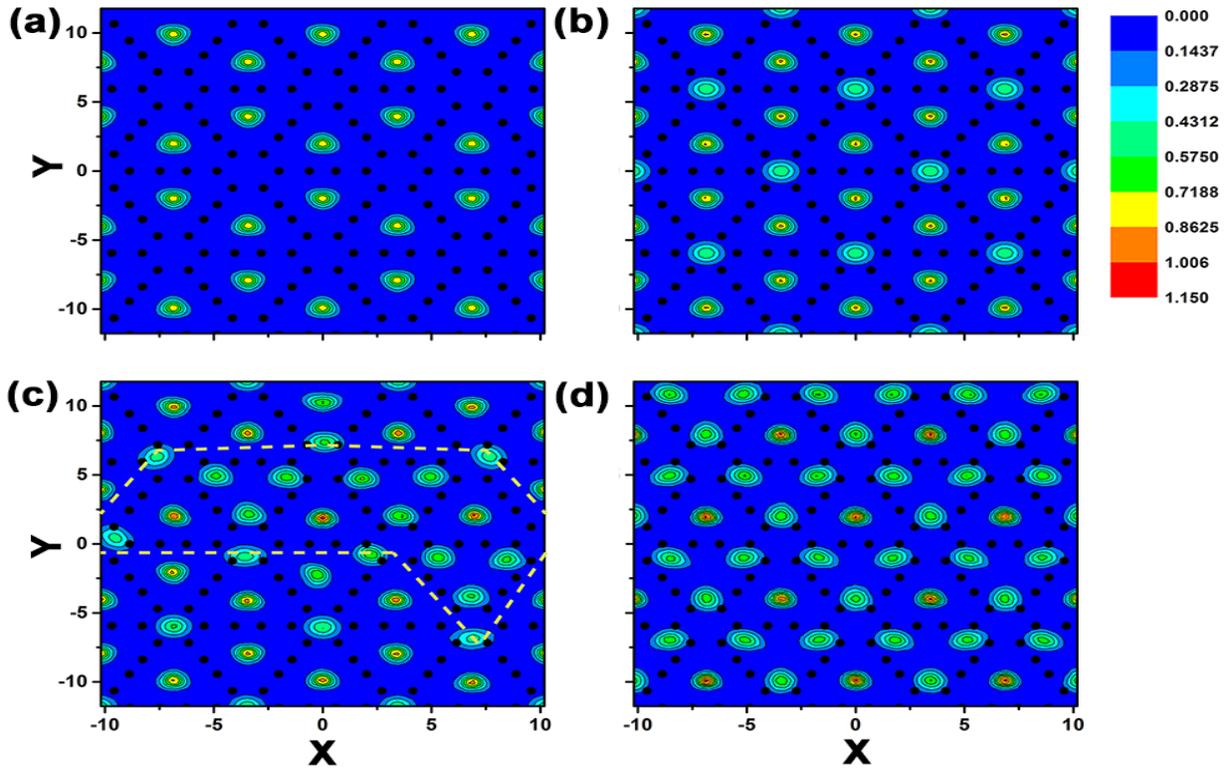}
\caption{(Color online) Two-dimensional density distributions of the first-layer $^4$He atoms
adsorbed on a single $\gamma$-graphyne sheet at areal densities of 
(a) 0.0491, (b) 0.0736, (c) 0.0859, and (d) 0.0982~\AA$^{-2}$.
The black dots represent the
positions of the carbon atoms of graphyne. The length unit is \AA~and
all contour plots are in the same color scale denoted by the color table in the upper right hand corner.
The yellow dotted lines in (c) separate two different domains from each other. 
The PIMC calculations were done at a temperature of $0.5$~K.
} 
\label{fig:2Dden}
\end{figure*}

For further analysis of different phases of the $^4$He monolayer,
we computed two-dimensional density distributions of $^4$He adatoms on $\gamma$-graphyne
at various areal densities.
In all four density plots presented in Fig.~\ref{fig:2Dden},
a distinct density peak represents the occupancy of a single first-layer $^4$He atom.
At an areal density of 0.0491~\AA$^{-2}$, which corresponds to the lowest energy state,
each of the irregular hexagons is seen in Fig.~\ref{fig:2Dden}(a) to accommodate one $^4$He atom 
at its center, confirming our conjecture made from the energetic analysis.
These $^4$He atoms form a honeycomb structure 
with the same primitive vectors 
as those of the underlying graphyne triangular lattice,
which is therefore a $1 \times 1$ registered phase in the Wood's notation.
It is also a C$_{2/3}$ commensurate solid with two out of every three adsorption sites
being occupied by $^4$He atoms. 
The lowest-energy state for the $^4$He monolayer on graphene 
is a C$_{1/3}$ commensurate solid~\cite{gordillo09}.
We note that the C$_{2/3}$ commensurate solid on graphyne is realized at an areal density significantly lower
than the C$_{1/3}$ commensurate helium coverage of 0.0636~\AA$^{-2}$ on graphene.
Furthermore, vacancies created in this C$_{2/3}$ solid on graphyne are found 
to be immobile and very weakly, if ever, interacting with each other, 
which could be understood by high potential barrier and long distances 
between the neighboring adsorption sites. 

As conjectured above, the local minima located at the centers of the small regular hexagons
accommodate additional $^4$He atoms beyond the C$_{2/3}$ commensurate coverage.
Figure~\ref{fig:2Dden}(b) shows another commensurate structure at an areal density of $\sigma=0.0736$~\AA$^{-2}$,
where each of the adsorption sites, both global minima and local minima, is occupied by a single $^4$He atom.
In this C$_{3/3}$ commensurate structure, 
the $^4$He adatoms form a triangular solid structure registered by $\frac{1}{\sqrt{3}} \times \frac{1}{\sqrt{3}}$ 
to the graphyne triangular lattice.
With further increase of the helium coverage beyond the C$_{3/3}$ solid,
where the energy per $^4$He atom increases monotonically as shown in Fig.~\ref{fig:energy},
the $^4$He monolayer enters a regime of various domain structures.
At higher $^4$He coverages, the $^4$He-$^4$He interaction as well as the $^4$He-substrate interaction 
is expected to affect the structure of the $^4$He monolayer.
At an areal density of 0.0859~\AA$^{-2}$, 
one can observe two different domains separated by the yellow dotted lines in Fig.~\ref{fig:2Dden}(c);
one domain involves some irregular hexagons accommodating three $^4$He atoms 
while the other consists of the $^4$He atoms in the C$_{3/3}$ commensurate order.
Another homogeneous phase of the $^4$He monolayer is observed 
at an areal density of $0.0982$~\AA$^{-2}$, where all $^4$He atoms are accommodated by irregular hexagons
and no small hexagon includes a $^4$He atom.
In this phase, some irregular hexagons accommodate three $^4$He atoms 
and the neighboring ones include only one $^4$He atom.
With an alternating order of the three-atom and the single-atom irregular hexagons,
the $^4$He atoms constitute another perfect triangular solid whose primitive vectors
are one half of those of the underlying graphyne structure.
This $\frac{1}{2} \times \frac{1}{2}$ registered phase is a C$_{4/3}$ commensurate solid
with 4 $^4$He atoms being accommodated by a graphyne unit cell.
We note that Li atoms attached to $\gamma$-graphyne 
could constitute an in-plane structure similar to this C$_{4/3}$ solid as reported in Ref.~\cite{hwang13} .
As discussed above, the $^4$He adatoms start to get promoted to the second layer
beyond the C$_{4/3}$ commensurate coverage of $0.0982$~\AA$^{-2}$. 
With further development of the second $^4$He layer, more $^4$He atoms 
are found to be squeezed into the first layer.
The fully-compressed first layer shows an incommensurate triangular lattice structure
like the corresponding layer on graphite. 

Our PIMC calculations have showed multiple distinct $^4$He layers on a single sheet of $\gamma$-graphyne 
which is not permeable to $^4$He atoms unlike $\alpha$-graphyne.
The $^4$He monolayer on $\gamma$-graphyne is found to exhibit various commensurate structures 
depending on the helium coverage,
including the $1 \times 1$, the $\frac{1}{\sqrt{3}} \times \frac{1}{\sqrt{3}}$, 
and the $\frac{1}{2} \times \frac{1}{2}$ registered phases. 
While some theoretical calculations predicted that zero-point vacancies would not be thermodynamically stable 
in bulk solid $^{4}$He~\cite{ceperley04,clark06,boninsegni06},
a substrate potential could stabilize the vacancy formation in a commensurate $^4$He solid on a substrate.
Therefore the existence of a stable commensurate structure is understood 
to be critical in realizing the vacancy-based supersolidity 
proposed originally by Andreev and Lifshitz~\cite{andreev69}.
One of several commensurate structures we found in the $^4$He monolayer on $\gamma$-graphyne
could manifest the superfluid response induced by vacancies, which is now under our investigation.

\begin{acknowledgments}
This work was supported by the Basic Science Research Program (2012R1A1A2006887)
through the National Research Foundation
of Korea funded by the Ministry of Education, Science and Technology.
We also acknowledge the support from the Supercomputing Center/Korea Institute 
of Science and Technology Information with supercomputing resources including technical support 
(KSC-2013-C3-033).
\end{acknowledgments}


\end{document}